\documentclass[spanish]{extarticle}
\usepackage[T1]{fontenc}
\usepackage[latin9]{inputenc}
\usepackage{units}
\usepackage{textcomp}
\usepackage{amsmath}
\usepackage{amsthm}
\usepackage{graphicx}
\usepackage[authoryear]{natbib}

\makeatletter
\numberwithin{equation}{section}
\numberwithin{figure}{section}
\newcommand{\lyxaddress}[1]{
	\par {\raggedright #1
	\vspace{1.4em}
	\noindent\par}
}

\@ifundefined{date}{}{\date{}}

\makeatother

\usepackage{babel}
\addto\shorthandsspanish{\spanishdeactivate{~<>}}

\begin{document}
\title{A cellular automaton model for thermal transport in low-dimensional
systems}
\author{Alejandra León Z.}
\maketitle

\lyxaddress{Instituto de Ciencias Básicas, Facultad de Ingeniería y Ciencia,
Universidad Diego Portales}
\begin{abstract}
In this work, we formulate a theoretical model based on a cellular
automaton (CA) to study thermal transport in low-dimensional nanostructures
across ballistic, diffusive, and transition regimes. Unlike computationally
intensive methods such as the Boltzmann Transport Equation (BTE),
our model stands out for its geometrical robustness, allowing the
seamless integration of substitutional impurities, vacancies, and
irregular edges. We validated the model using graphene nanoribbons
(AGNRs), successfully replicating the dependence of thermal conductivity
on ribbon width and temperature. Results demonstrate that the model
captures critical scattering and confinement effects with a linear
scalability $O(N)$. Given the increasing pressure to optimize computational
resources and reduce the carbon footprint associated with AI infrastructure,
this CA model emerges as a highly efficient tool for the parametric
exploration and design of next-generation thermal devices.
\end{abstract}

\section*{Introduction:}

The rapid growth of artificial intelligence (AI) and its associated
infrastructures has significantly intensified the global energy demand
of data centers, raising increasing concerns regarding carbon footprint,
water consumption, and stress on electrical grids. Recent studies
indicate that large-scale AI training and inference already account
for a growing fraction of worldwide electricity consumption, with
projections pointing to a substantial increase over the next decade
if no improvements in computational and energy efficiency are introduced
{[}1--3{]}. In this context, the scientific community has begun to
emphasize the urgent need for numerical modeling tools that, in addition
to being physically consistent, exhibit reduced computational cost,
thereby enabling extensive parametric explorations without relying
on intensive supercomputing resources {[}1,2{]}.

This challenge is particularly relevant in the study of nanoscale
heat transport, where the most rigorous available theoretical approaches
are also among the most computationally demanding. Consequently, there
is a growing motivation to develop and validate simplified models
capable of capturing the dominant mechanisms of phonon-mediated thermal
transport, while remaining suitable for exploratory studies, complex
geometrical configurations, and systematic parameter sweeps, all within
a low computational footprint compatible with more sustainable research
practices {[}3--5{]}.

The most comprehensive theoretical framework for describing phonon-mediated
heat transport in crystalline materials is the phonon Boltzmann transport
equation (BTE). This approach provides a unified description of ballistic,
diffusive, and intermediate regimes, incorporating spectral, polarization,
and directional dependencies of phonons. In recent years, numerous
studies have highlighted that, in nanostructures and two-dimensional
materials, thermal transport deviates significantly from Fourier\textquoteright s
law, requiring BTE-based descriptions to properly capture confinement
effects, scattering mechanisms, and nonequilibrium behavior {[}6--8{]}.

However, the high dimensionality of the BTE---spanning real space,
reciprocal space, time, and multiple phonon modes---renders its direct
numerical solution computationally demanding. Even modern implementations
based on deterministic or Monte Carlo schemes exhibit rapidly increasing
computational costs when complex geometries, defects, irregular boundaries,
or transient phenomena are considered {[}6,9{]}. This has motivated
the development of parallel and GPU-accelerated solvers, which can
significantly reduce computation times but still require advanced
computational infrastructures and sophisticated numerical implementations
{[}10{]}.

As an intermediate alternative between the fidelity of full BTE solvers
and the simplicity of macroscopic models, mesoscopic discrete methods
have been explored, including lattice-based approaches such as the
lattice Boltzmann method (LBM) applied to thermal and phonon transport.
These methods are inspired by kinetic discretizations of the BTE and
are characterized by local propagation and collision rules on discrete
spatial lattices, which facilitate parallelization and reduce computational
cost compared to full BTE solvers {[}11,12{]}. Several studies have
shown that lattice-based approaches can reproduce general trends in
thermal transport and capture geometry- and boundary-induced effects,
although they often require system-specific parametrizations and do
not always preserve a direct microscopic physical interpretation {[}11{]}.

In parallel, an active line of research has emerged combining physical
models with machine learning techniques to accelerate BTE solutions
or to construct surrogate models trained on high-fidelity data. In
particular, physics-informed neural networks (PINNs) and hybrid Monte
Carlo--machine learning methods have demonstrated the ability to
approximate multiscale phonon transport with a significant reduction
in computational time during the inference stage {[}13--15{]}. Despite
their potential, these approaches rely on datasets generated using
computationally expensive methods and may lose robustness outside
their training domain, which limits their direct applicability as
general exploratory tools. As a result, they continue to coexist with
simplified mechanistic models that do not require prior training and
retain a direct physical interpretation of their evolution rules {[}14,15{]}.

Overall, the state of the art indicates that phonon-mediated thermal
transport in nanostructures is an intrinsically multiscale and geometry-dependent
problem, for which the most rigorous modeling techniques are associated
with high computational costs {[}6--10{]}. The increasing energy
demand of large-scale computation, further exacerbated by the expansion
of AI technologies, reinforces the need for alternative modeling strategies
that are simple, interpretable, and computationally efficient, enabling
systematic and exploratory studies with a reduced environmental footprint
{[}1--3{]}.

Within this context, cellular automaton--based models emerge as a
complementary tool to existing approaches. They allow effective descriptions
of phonon transmission, scattering, and confinement mechanisms through
local discrete rules, naturally accommodate complex geometries, defects,
and irregular boundaries, and enable extensive parametric studies
at minimal computational cost. Such models can therefore be regarded
as a first-level analytical framework, either prior to or in conjunction
with more expensive BTE-based or atomistic simulations, while aligning
with current demands for more efficient and sustainable computational
practices. In this work, we formulate a cellular automaton model to
simulate phonon-mediated thermal transport at the atomic scale in
low-dimensional systems, with particular emphasis on capturing the
effects of vacancies, impurities, irregular edges, disorder, and complex
geometries through an effective, coarse-grained description.

\section*{Model Description:}

A cellular automaton (CA) consists of a discrete spatial array of
cells, each characterized by a finite set of state variables whose
temporal evolution is determined by local rules depending on the neighborhood.
Despite the simplicity of these rules, local interactions can give
rise to complex collective dynamics, making cellular automata a particularly
suitable tool for studying complex physical systems and emergent phenomena.
This approach has been widely used in physics to provide effective
descriptions of multiscale processes with reduced computational cost
{[}16-21{]}. The version presented in this work is optimized for one-dimensional
and two-dimensional systems, although its extension to three dimensions
is straightforward.In the proposed model, each cell of the automaton
corresponds to an atomic site of the nanostructure under study. It
is important to emphasize that, since phonons are collective excitations
of the crystal lattice, they cannot be strictly assigned to an individual
atom. Consequently, the present model adopts an effective coarse-grained
description, in which the local variables associated with each cell
do not represent inherent phonon modes, but rather local contributions
to the collective vibrational excitations of the system.The state
of each cell \$i\$ is defined by the following vector:

\begin{equation}
S_{i}\left(t\right)=\left(i,x,y,N_{a},N_{o},N_{f},N_{T},T\right)
\end{equation}

where $i$ identifies the atomic species present at the site (e.g.,
carbon, nitrogen, silicon, or a vacancy), and $x$ and $y$ correspond
to the spatial coordinates of the atomic site for two-dimensional
systems. These components remain fixed throughout the simulation and
define the geometry and composition of the system.The remaining components
of the state vector evolve over time and describe the thermal dynamics
of the system. The variables $N_{a}$, $N_{o}$, and $N_{f}$ represent
effective occupations associated with acoustic, optical, and flexural
phonons, respectively. These quantities do not correspond to phonons
localized at an atomic site, but rather to a local projection of the
collective vibrational energy associated with each type of excitation
within the adopted mesoscopic description. The variable $N_{T}$ denotes
the total effective number of vibrational excitations present in the
cell, while $T$ represents the local temperature, defined based on
the effective vibrational energy contained at the site The initialization
of the vibrational fields is performed by assigning to each atomic
site i i (carbon, nitrogen, or vacancy) effective occupations $(N_{ai},N_{oi},N_{fi})$
consistent with the prescribed initial temperature profile. Within
the coarse-grained formulation adopted here, these variables represent
local projections of the collective vibrational energy rather than
microscopic Bose--Einstein phonon populations. The total effective
occupation is set through a calibrated mapping $T_{i}=\alpha N_{Ti}$,
where $N_{Ti}=N_{ai}+N_{o,}+N_{fi}$ and $\alpha\simeq4.8$ defines
the internal temperature scale of the cellular automaton. This calibration
factor was chosen to ensure that benchmark simulations of pristine
armchair graphene nanoribbons reproduce the correct order of magnitude
and temperature dependence of the thermal conductivity reported in
the literature. The branch-resolved initial occupations are then obtained
by distributing $N_{Ti}$ among acoustic, optical, and flexural channels
using fixed, species-dependent weighting factors $w_{a,o,f}^{(s)}$
, with $w_{a}^{(s)}+w_{o}^{(s)}+w_{f}^{(s)}=1$, after which the occupations
evolve according to the local update rules of the cellular automaton.

Under this approach, the cellular automaton does not attempt to explicitly
resolve the phonon spectrum of the system or the dynamics of individual
normal modes. Instead, it seeks to effectively capture the dominant
mechanisms of transmission, scattering, and confinement of vibrational
energy through discrete local rules. This allows for the study of
thermal transport in nanostructures with complex geometries, defects,
impurities, and irregular edges, while maintaining a reduced computational
cost. The temporal evolution of the cellular automaton is implemented
through a local update rule that acts on the effective phonon occupations
of each cell. At each time step, the dynamic variables associated
with an atomic site are modified based on the exchange of vibrational
energy with its nearest neighbors, preserving the local and discrete
nature of the model. The intensity of this exchange is controlled
by the \$\textbackslash beta\$ parameter, defined as an effective
phonon coupling parameter, which regulates the fraction of vibrational
energy transferred between adjacent cells during each update. This
rule does not aim to explicitly reproduce a specific microscopic phonon
scattering mechanism, but rather to effectively capture the local
redistribution of energy associated with thermal transport. By adjusting
the value of \$\textbackslash beta\$, the model allows for a continuous
interpolation between behaviors dominated by ballistic-type transport
and regimes characterized by diffusive dynamics, maintaining at all
times a low computational cost and a clear physical interpretation
at the mesoscopic level. The update function for the number of effective
occupations $N$ in the acoustic, optical, or flexural modes, at site
$i$, is

\begin{equation}
N_{i}\left(t+1\right)=N_{i}\left(t\right)+\beta\left(\sum_{j\:\epsilon\:n\left(i\right)}N_{j}\left(t\right)-z_{i}N\left(t\right)\right)
\end{equation}

In this expression, $n(i)$ represents the set of nearest neighbors
of site $i$, and $z_{i}=|n(i)|$ is its local coordination number.
The parameter $\beta$ follows a numerical modeling mediated by a
sigmoid function of the type:

\begin{equation}
\sigma\left(T_{m}\right)=\left(1+e^{\frac{T_{m}-T_{0}}{\triangle T}}\right)^{-1}
\end{equation}

The characteristic temperature $T_{0}$ represents the approximate
temperature that separates the ballistic from the diffusive regime
for the considered atom, while the parameter $\Delta T$ controls
the smoothness of this transition. The results obtained in the validation
simulations proved to be robust against moderate variations of these
parameters. Thus, the beta parameter can be written as $\beta(T_{m})=\beta_{0}\sigma(T_{m})$.
In this expression, $T_{m}$ represents the average temperature of
the system, and $\beta_{0}$ depends on the atom at the site being
updated. The coupling parameters $\beta_{0}$ associated with the
different phonon branches are effective mesoscopic magnitudes that
control the local exchange of vibrational energy between neighboring
sites. Their values are not derived from microscopic material constants
but are selected within physically reasonable ranges to ensure numerical
stability and reproduce qualitatively correct transport regimes. Similar
effective parametrizations are commonly adopted in mesoscopic and
lattice-based transport models {[}11, 12, 16{]}. The specific values
of $\beta_{0}$ were initially selected based on physical considerations
regarding the relative mobility of acoustic, optical, and flexural
modes, and were subsequently refined through calibration against reference
thermal transport behaviors.

\begin{equation}
k=\frac{J}{\nicefrac{dT}{dx}}
\end{equation}

where $k$ represents the thermal conductivity of the system, $J$
denotes the heat energy per unit time and unit area, and the denominator
$dT/dx$ represents the thermal gradient. The time unit is determined
using the time step of the automaton's evolution, the atomic bond
lengths, and the propagation velocity of the considered oscillation
modes. Finally, the implementation of the model in a computational
code allows for real-time mapping of local vibrational excitations
in the optical, acoustic, and flexural ranges, and from this mapping,
the local temperature of the system is obtained.

\section*{Model Validation and Proof of Concept:}

\subsection*{Phonon Channels and Scattering Effects.}

Model validation was performed using graphene nanoribbons (AGNRs),
given the extensive body of theoretical and experimental studies available
for these systems, which allow for a preliminary validation of the
model. The first system studied is a small, pristine graphene nanoribbon
with a length of 110 Å and armchair edges. Figure 1 shows the dependence
of thermal conductivity as a function of the ribbon width, measured
in units of carbon dimer lines, for three different temperatures.
The number of lines expressed as length, for Figure 1, ranges from
13 Å to 62 Å.We can observe an increase in thermal conductivity with
AGNR width and a decrease in conductivity with temperature. This is
consistent with theoretical works based on other frameworks {[}22-23{]}.
The increase in \$\textbackslash kappa\$ with ribbon width, due to
the higher number of phonon channels available for transmission, is
captured by our model because the available update cells in the direction
transverse to the heat flow increase with width. The second behavior
is due to scattering and Umklapp effects, which our model recovers
through the relationship between the beta coefficient and temperature.
The temperature shown in the graph corresponds to the mean temperature
$\left(T_{m}\right)$ of the system, and the boundaries are simulated
with temperatures $T_{H}=T_{m}+10\;K$ y $T_{C}=T_{m}-10\;K$.
\begin{center}
\begin{figure}
\begin{centering}
\includegraphics[scale=0.3]{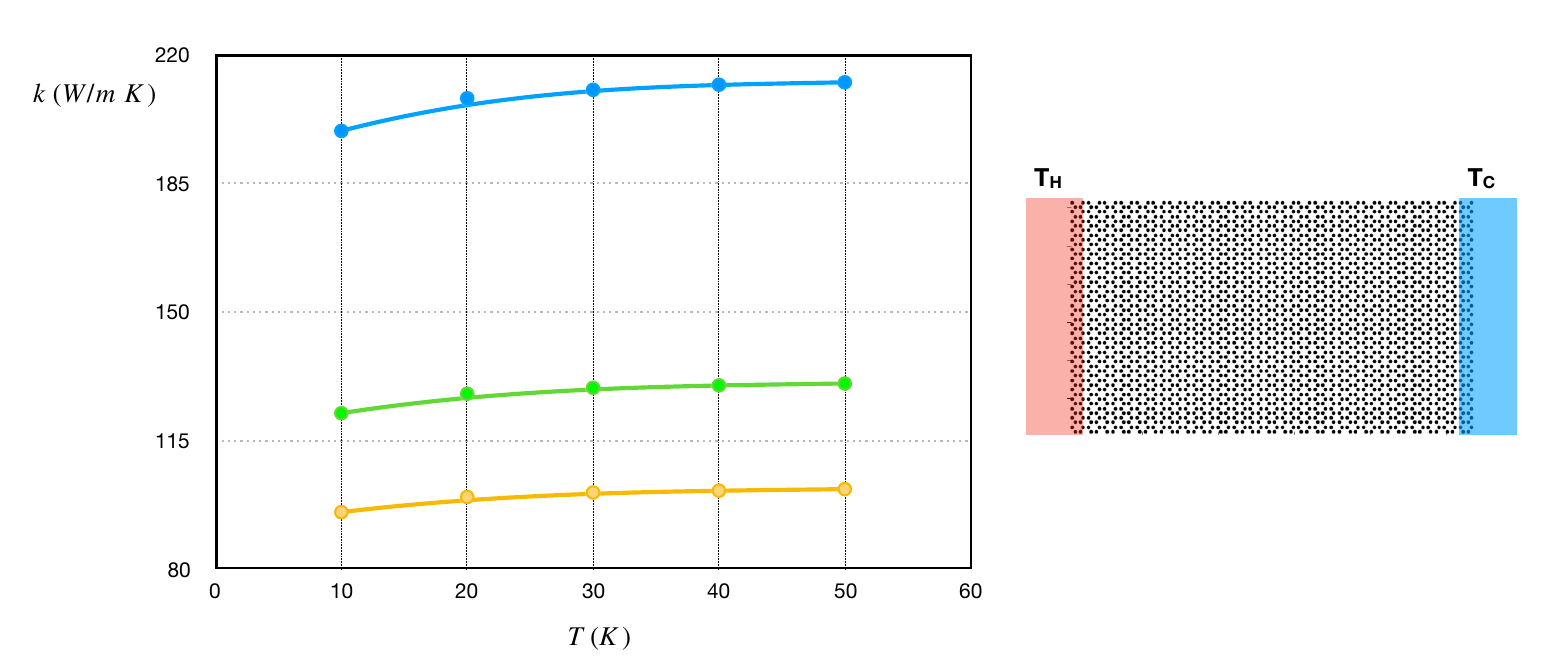}
\par\end{centering}
\caption{Thermal conductivity of pristine AGNRs as a function of width and
temperature. The figure also includes a schematic of the physical
system, including the thermal reservoirs.}

\end{figure}
\par\end{center}

\subsection*{Inclusion of Vacancies and Irregular Edges.}

The main strength of this model lies in its potential to study disordered
systems and non-conventional geometries, allowing for theoretical
proof-of-concept tests to investigate, in a first approximation, the
heat flow dynamics in these systems and, consequently, perform trial
engineering for thermal device design. The second case studied with
the model consisted of a pristine AGNR of fixed width as a function
of temperature, which was contrasted with the same structure but containing
1\% randomly arranged vacancies. Additionally, rough edges with a
5\% penetration level were considered. Since both the edges and the
vacancy positions are randomly distributed, a statistical study was
conducted, reporting the arithmetic mean and standard deviation for
each case. These results are shown in Figure 2.

\begin{figure}
\begin{centering}
\includegraphics[scale=0.3]{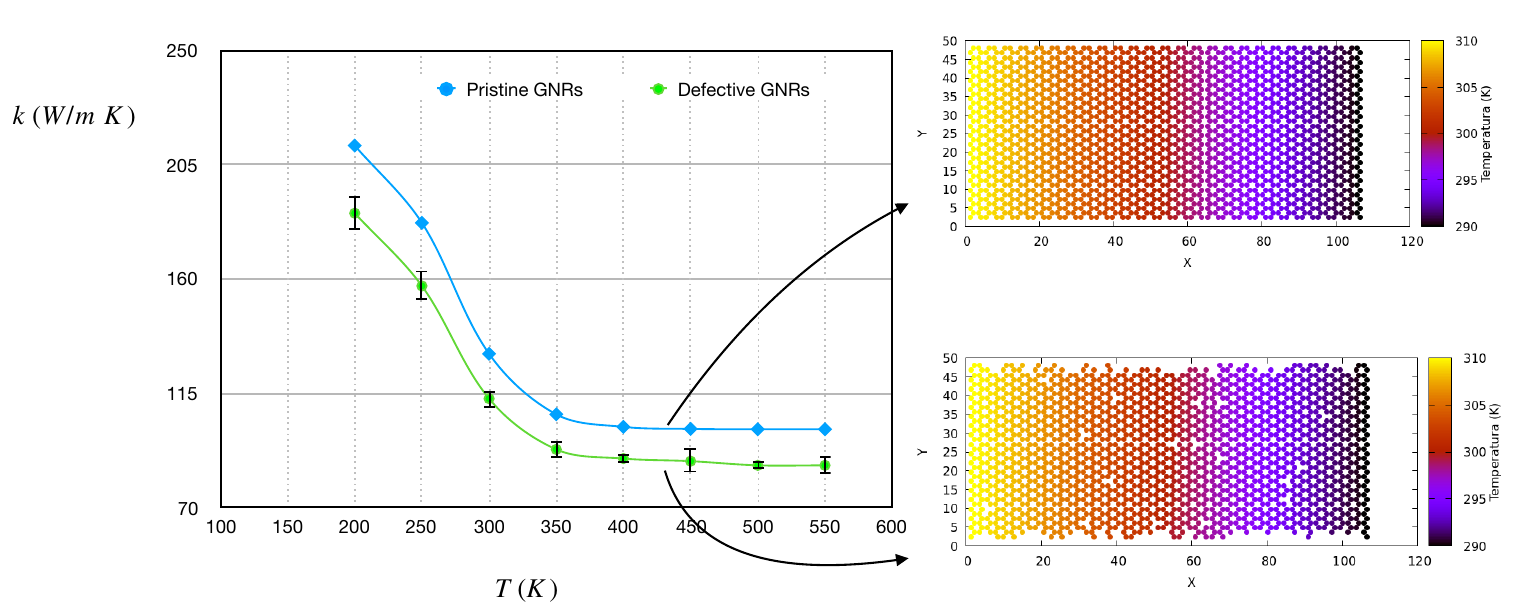}
\par\end{centering}
\caption{Thermal conductivity as a function of temperature for an AGNR with
a length of 110 Å and a width of 40 lines, corresponding to 50 Å.
The figure also displays the results for the system with vacancies
and irregular edges. Additionally, temperature maps for both cases
are included.}

\end{figure}

We can observe that thermal conductivity decreases with vacancies
and irregular edges across all studied temperatures. This behavior
is consistent with disorder studies utilizing different research platforms
{[}24{]}. The study can incorporate maps showing the local contribution
to the system's thermal oscillations in acoustic, optical, and flexural
modes, or a system temperature mapping, such as the one shown in Figure
2.

\subsection*{Engineering the geometry of the studied systems}

An additional test was performed to further investigate the behavior
of the model, consisting of the analysis of an S-shaped nanoribbon
segment featuring two 90° bends. In this system, a clear discontinuity
in the temperature gradient along the structure was observed. Consequently,
we defined an effective thermal conductivity, $k_{\mathrm{eff}}$,
by using both ends of the system to compute the temperature gradient.

Figure 3 shows the effective thermal conductivity as a function of
the length of the central region of the structure. To this end, we
introduced the coefficient $\mathrm{Crf}$ (central region factor),
defined as the fraction---measured along the $x$-axis---of the
distance between the hot and cold reservoirs assigned to the two arms
of the S-shaped geometry. For example, $\mathrm{Crf}=3$ corresponds
to three segments of equal length along the x x-direction. A value
of $\mathrm{Crf}=4$ implies that the arms have a length equal to
one quarter of the total length, while the central region spans one
half of the total length. Fractional values of $\mathrm{Crf}$ were
also considered in order to progressively reduce the size of the central
region, reaching values as low as 2.1.

In the same Figure 3, temperature maps corresponding to several S-shaped
systems are presented. From these results, we infer that the central
region effectively behaves as a semi-open thermal bottleneck, whose
transmissivity to heat progressively decreases as its length is reduced.

\begin{figure}
\begin{centering}
\includegraphics[scale=0.3]{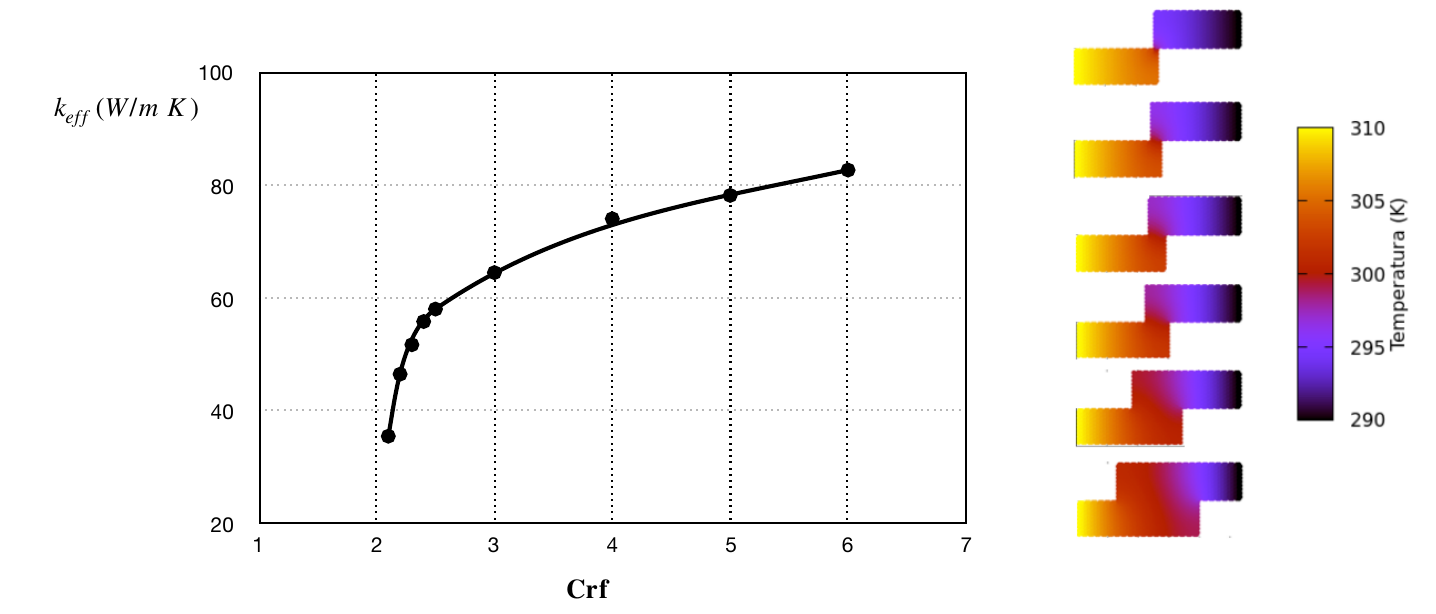}
\par\end{centering}
\caption{Effective thermal conductivity as a function of the length of the
central region of the structure. Temperature maps are also included
for six representative values out of the nine shown in the thermal
conductivity plot: $Crf=2.1,\;2.2,\;2.3,\;2.4,\;3.0\;y\;4.0$.}

\end{figure}

The effective thermal conductivity decreases rapidly as the $\mathrm{Crf}$
factor approaches 2. A particularly interesting outcome of this part
of the study is the emergence of a discontinuity in the thermal gradient
across the system. This jump becomes increasingly pronounced as the
factor tends toward this limiting value. This stage of the validation
demonstrates that the model naturally evolves toward well-established
physical scenarios associated with energy transport. When the factor
is close to this limit, the arms of the structure acquire the temperatures
of the thermal reservoirs, effectively resulting in two physical systems
that are almost thermally disconnected, as illustrated in Fig. 4 for
two representative cases. This behavior suggests potential applications
in thermal rectification or in the design of structurally induced
topological thermal insulators.

\begin{figure}
\begin{centering}
\includegraphics[scale=0.3]{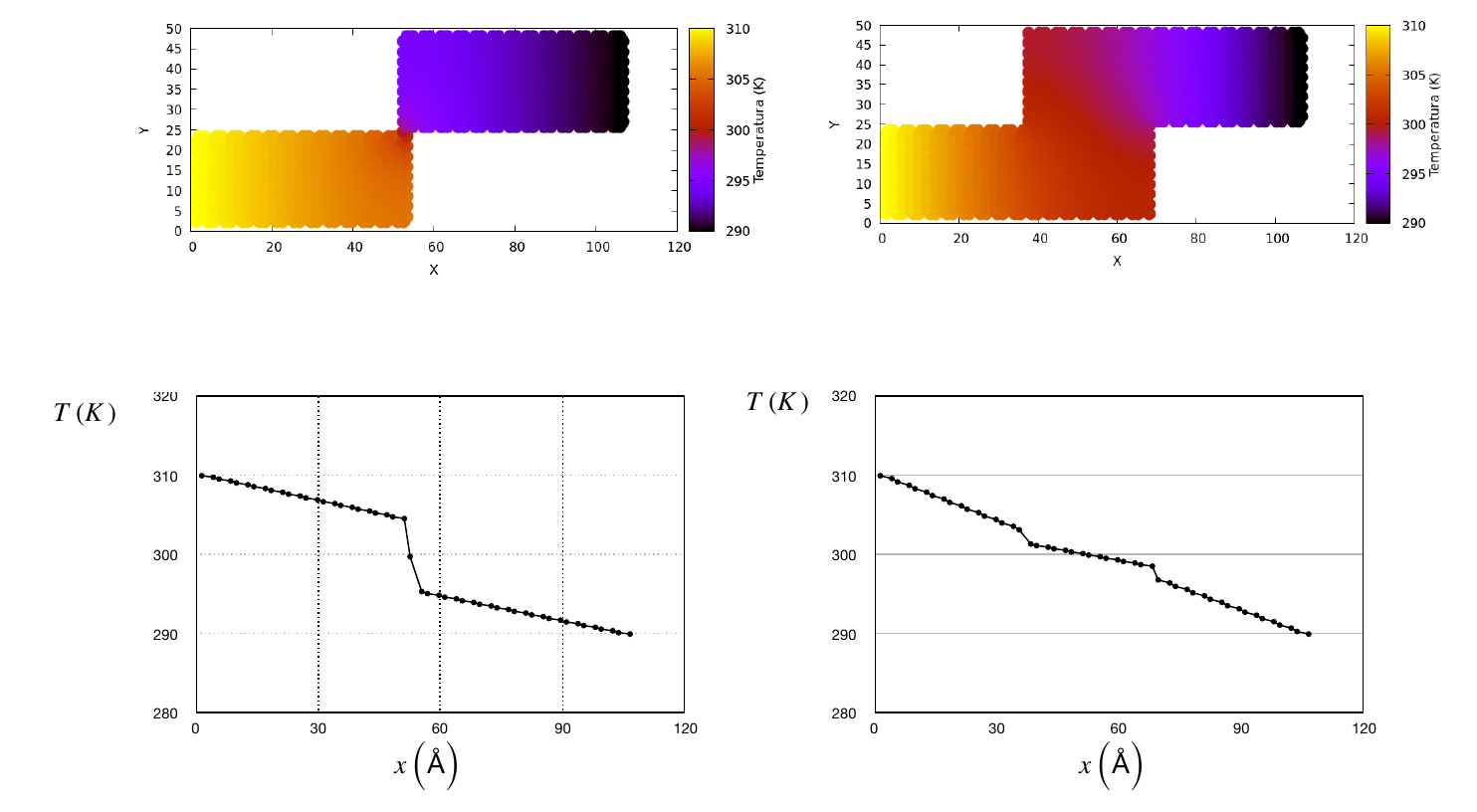}
\par\end{centering}
\caption{Thermal gradient and temperature maps for two S-shaped structures.
The two panels on the left correspond to a factor $\mathrm{Crf}=2.1$,
while the panels on the right correspond to $\mathrm{Crf}=3.0$.}
\end{figure}

\subsection*{Scalability and computational methodology}

The cellular automaton model was developed using the Lazarus IDE v3.2
(Free Pascal), enabling efficient native execution without the heavy
dependencies commonly associated with other computational environments.
The resulting application integrates five real-time graphical visualization
components: the nanoribbon temperature map, the thermal gradient,
the thermal conductivity, the incoming and outgoing heat fluxes, and
the average temperature of the sample.

For this reason, the scalability study---performed in a Linux environment
(Ubuntu 22.04.5 LTS) on an HP ProBook 440 G7 equipped with an x86-64
architecture---reports execution times that comprehensively account
for both the algorithmic computational processing and the computational
cost associated with synchronous graphical rendering.

\begin{figure}
\begin{centering}
\includegraphics[scale=0.4]{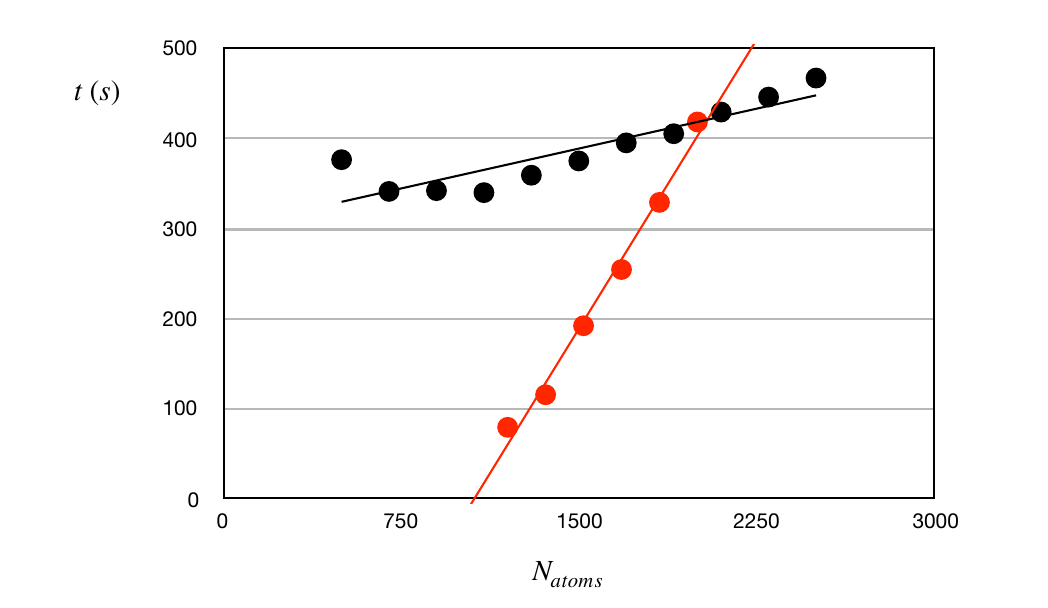}
\par\end{centering}
\caption{Model scalability study. The graph displays the execution time in
seconds as a function of the total number of atoms in the simulated
system. The dark points represent transversal scalability (increasing
nanoribbon width with constant length), while the light points represent
longitudinal scalability (increasing length with constant width).
In both studies, 1\% vacancies and irregular edges with 5\% penetration
were considered.}

\end{figure}

For cellular automaton models that simulate phenomena governed by
strictly local interactions, the execution time is expected to scale
linearly with system size {[}16--21{]}. This behavior arises because
updating each cell requires the same computational effort and, since
all cells are updated in parallel, the total update time scales linearly.
Nevertheless, this scaling behavior was explicitly verified in the
present work, as the model is designed to simulate structures that
may grow either longitudinally or transversely with respect to the
direction of the thermal flux, which in turn modifies the scalability
slope.

Figure 5 presents the results of the scalability study performed for
graphene nanoribbons. The total execution time of the global program---measured
in seconds---is reported. As mentioned above, the program incorporates
real-time graphical visualization, with on-screen plots updated every
1000 cellular automaton iterations. In the transverse scalability
study, a slight initial decrease in execution time is observed, followed
by a linear increase. This initial drop can be attributed to the fact
that thin nanoribbons with irregular edges and vacancies require a
longer time to reach the steady-state regime compared to wider ribbons.
The difference in the slopes observed in the two scalability analyses
is explained by the convergence dynamics of the cellular automaton
toward the steady state, namely that, on average, heat flows longitudinally
along the nanoribbon.

\section*{Conclusions and Future Work:}

The model presented in this study provides a reliable and computationally
efficient mechanistic approach to thermal transport mediated by phonons
in irregular and disordered nanostructures. Its primary novelty lies
in its pure CA formulation, which avoids the direct simulation of
particles, instead inferring heat flow and global thermal properties
through local update rules. The model accurately reproduces the thermal
behavior of graphene nanoribbons reported in previous literature,
capturing the essential physics of phonon scattering and Umklapp processes.
Furthermore, the $O(N)$ scalability, combined with the flexibility
to incorporate complex geometries---such as the S-shaped ribbons
studied here---proves that this tool is not only effective but sustainable
for large-scale parametric sweeps. This is particularly relevant in
the current era of 'green computing,' where reducing the energy cost
of numerical simulations is paramount. Future work will focus on applying
this model to study the thermal performance of novel 2D materials
and complex van der Waals heterostructures, which will be published
in upcoming reports.

\section*{Acknowledgments:}

This research was carried out with the support of the Fondecyt Regular
Project \#1211913 and the Faculty of Engineering and Sciences of the
Diego Portales University.

\section*{References:}

1. de Vries, A. The growing energy footprint of artificial intelligence.
Joule (2023). 

2. de Vries-Gao, A. The carbon and water footprints of data centers
and what this could mean for artificial intelligence. Patterns (2025). 

3. de Vries-Gao, A. Artificial intelligence: supply chain constraints
and energy implications. Joule (2025). 

4. Beardo, A. et al. Nanoscale confinement of phonon flow and heat
transport. npj Computational Materials (2025). 

5. Kim, K., Liu, W. Molecular dynamics simulations in nanoscale heat
transfer: A mini review. Journal of Heat Transfer (2025). 

6. Li, Z. et al. Phonon Boltzmann transport equation based modeling
of heat conduction: progress and challenges. Journal of Applied Physics
(2022). 

7. Cepellotti, A., Marzari, N. Transport by phonons in two-dimensional
materials. Nature Reviews Materials (2016). 

8. Lee, S. et al. Hydrodynamic phonon transport in low-dimensional
materials. Nature Communications (2015). 

9. Minnich, A. Advances in phonon Monte Carlo methods. Journal of
Physics: Condensed Matter (2015). 

10. Tran, H. D. et al. Parallel computation of the phonon Boltzmann
transport equation. SIAM Journal on Scientific Computing (2025). 

11. Guo, Z. Lattice Boltzmann modeling of phonon transport. Journal
of Computational Physics (2016). 

12. An, B. Grid technologies in the lattice Boltzmann method: review
and advances. Mathematics (2025). 

13. Li, R. et al. Physics-informed deep learning for solving the phonon
Boltzmann transport equation. npj Computational Materials (2022). 

14. Lin, Q. et al. Monte Carlo physics-informed neural networks for
multiscale heat conduction. Journal of Computational Physics (2025). 

15. Wang, Y. et al. Machine learning accelerated phonon transport
modeling. Computational Materials Science (2024).

16. B. Chopard and M. Droz. Cellular automata modeling of physical
systems. Cambridge University Press (2005).

17. A. León, Z. Barticevic, M. Pacheco, Solid State Commun. 152 (2012)
41--44.

18. A. León, Curr. Appl. Phys. 13 (2013) 2014--2018.

19. A. León and M. Pacheco, Phys. Lett. A 375, 4190 (2011).

20. A. León, Comput. Phys. Commun. 183 (2012) 10.

21. A. León, J. Magn. Magn. Mater. 340 (2013) 120--126.

22. Zhixin Guo; Dier Zhang; Xin-Gao Gong. Appl. Phys. Lett. 95, 163103
(2009).

23. Haiying Yang et al. J Mol Model (2013) 19:4781--4788.

24. Evans, W. J., et al. (2010). Effect of size and edge structure
on the thermal conductivity of graphene nanoribbons. Physical Review
B, 82(21), 214301.
\end{document}